\documentclass[10pt, conference, letterpaper]{IEEEtran}

\usepackage{graphicx}
\usepackage{subfig}
\usepackage{enumerate}
\usepackage{amsmath}
\usepackage{cases}
\usepackage{amsfonts}
\usepackage{algorithm}
\usepackage{algorithmic}
\usepackage{flushend}
\usepackage{multirow}
\usepackage[
  pass,% keep layout unchanged
]{geometry}
\begin{document}

\title{On Crowdsourced Interactive Live Streaming: A Twitch.TV-Based Measurement Study}

\author{\IEEEauthorblockN{Cong Zhang, Jiangchuan Liu}
\IEEEauthorblockA{School of Computing Science\\
Simon Fraser University, Burnaby, BC, Canada\\
Email: \{congz, jcliu\}@cs.sfu.ca}
}

\maketitle
\begin{abstract}
Empowered by today's rich tools for media generation and collaborative production, the multimedia service paradigm is shifting from the conventional single source, to multi-source, to many sources, and now toward {\em crowdsource}. Such crowdsourced live streaming platforms as Twitch.tv allow general users to broadcast their content to massive viewers, thereby greatly expanding the content and user bases. The resources available for these non-professional broadcasters however are limited and unstable, which potentially impair the streaming quality and viewers' experience. The diverse live interactions among the broadcasters and viewers can further aggravate the problem.

In this paper, we present an initial investigation on the modern crowdsourced live streaming systems. Taking Twitch as a representative, we outline their inside architecture using both crawled data and captured traffic of local broadcasters/viewers. Closely examining the access data collected in a two-month period, we reveal that the view patterns are determined by both events and broadcasters' sources. Our measurements explore the unique source- and event-driven views, showing that the current delay strategy on the viewer's side substantially impacts the viewers' interactive experience, and there is significant disparity between the long broadcast latency and the short live messaging latency. On the broadcaster's side, the dynamic uploading capacity is a critical challenge, which noticeably affects the smoothness of live streaming for viewers.

 \end{abstract}

\section{Introduction}

Empowered by today's rich tools for media generation and collaborative production, the multimedia service paradigm is shifting from the conventional single source, to multi-source, to many sources, and now toward {\em crowdsource}~\cite{Simoens:2013:MobiSys}, where the available media sources for the content of interest become highly diverse and scalable. In the 2014 Sochi Winter Olympics, NBC (National Broadcasting Company) had a total of 41 live feeds distributed both in Sochi and in USA, and in the 2014 FIFA World Cup, when a goal is scored, CBC (Canadian Broadcasting Corporation) synchronized the live scenes of the cheering fans in public squares from cities worldwide in its live streaming channel. The evolution is driven further by the advances in personal and mobile devices that can readily capture high quality audio/video anywhere and anytime (e.g., iPhone 6 supports 60 fps 1080p High Definition (HD) video recording, and 240 fps slow-motion recording for 720p HD videos).

Crowdsourced content creation is expected to usher in a new wave of innovations in how multimedia content is created and consumed, allowing content creators from different backgrounds, talents, and skills to collaborate on producing future multimedia content. For instance, the industrial pioneer, Twitch.tv (www.twitch.tv), allows anyone to broadcast their content to massive viewers, and the primary sources come from game players from PCs or other gaming consoles, e.g., PS4 and XBox. According to Twitch's Retrospective Report 2013\footnote{http://www.twitch.tv/year/2013}, in just three years, the number of viewers grew from 20 million to 45 million, while the number of unique broadcasters tripled to 900 thousand. Other similar platforms such as Poptent (www.poptent.com) and VeedMe (www.veed.me) have emerged in the market with great success, too.

Existing works have identified the characteristics of popular streaming systems~\cite{Xu:2013:YouTube, Li:2012:IMC:PPLive, Hwang:2013:ICNP, Zhou:2012:NOSSDAV}.
In contrast to these conventional streaming systems, e.g., YouTube-like streaming, mobile live streaming, and P2P live streaming, Twitch-like crowdsourced interactive live streaming has a number of distinguished features. First, Twitch-like services do not provide the sources of live streaming by themselves. Rather, they serve as a platform that bridges sources and viewers, thereby greatly expanding the content and user bases. On the other hand, the sources are no longer professional content producers and providers, and often have limited computation and network resources. They can even join or leave at will, or crash at anytime. All these make high-quality live streaming more challenging. Second, Twitch-like services also promote viewers' involvement with live content broadcasters.
The viewers can choose their preferred perspective for a live event (e.g., one particular game player, or a game commentator) and enjoy virtual face-to-face interactions with real-time chatting. It is necessary to ensure timely interaction and minimize the switching latencies, which again is aggravated with the multiple non-professional sources and the massive viewers.

In this paper, we present an initial investigation on the modern crowdsourced live streaming systems. Taking Twitch as a representative, we outline their inside architecture using both crawled data and captured traffic of local broadcasters/viewers. Closely examining the access data collected in a two-month period, we reveal that the view patterns are determined by both events and broadcasters' sources. Our
measurements explore the unique source-driven and event-driven views, showing that the current delay strategy on the viewer's side substantially impacts the viewers' interactive experience, and there is significant disparity among the long broadcast latency and the short live messaging latency. On the broadcaster's side, the dynamic uploading capacity is a critical challenge, which noticeably affects the smoothness of live streaming for viewers. Inspired by the measurement results, we discuss potential enhancements toward better crowdsourced interactive live streaming.

The remainder of this paper is organized as follows.
Section~\ref{sec:inside} introduces our measurement methodology, which sheds insight into the architecture of Twitch.
Section~\ref{sec:viewfeatures} details the views pattern in Twitch.
We present the latency results on the viewer's side and
analyze the impacts of sources on the broadcaster's side in Section~\ref{sec:viewer}.
Finally, Section~\ref{sec:conclusion&further} concludes the paper with further discussions.

\section{Inside the Twitch Architecture}
\label{sec:inside}
\begin{table*}[!th]
\small
\center
\caption{The details about Twitch REST APIs in our crawler}
\begin{tabular}{|l|l|}
\hline
\textbf{REST APIs}            & \textbf{Description}                                      \\ \hline
GET /streams/summary          & Get the global statistics of streams and views at present \\ \hline
GET /streams                  & Get the meta file of live streams at present         \\ \hline
GET /channels/:channel        & Get the number of total views, followers and delay setting of broadcaster's channel    \\ \hline
GET /channels/:channel/videos & Get the number of total views, duration of each stream in broadcaster's channel \\ \hline
\end{tabular}
\label{tab:api}
\end{table*}
%XXX
As a new generation and proprietary system, despite certain information leakages~\cite{Twitch:justin}, the inside details of Twitch and particularly the access data remain unclear to the public,  so do many other crowdsourced live streaming systems in the market. With the assistance of Twitch's Representational State Transfer (REST) APIs\footnote{http://dev.twitch.tv/}, we continually crawled the access data of live contents from Twitch in a two-month period (from October 1st to November 30th, 2014). The crawled data include the number of Twitch streams, the number of Twitch views, and the meta-data of live streams every ten minutes. The meta-data include that the game name, stream ID, broadcaster's channel ID, current views, created time and other information. Our crawler analyzed these meta files to create the sets of broadcasters' channels and scrape the number of the total views and durations of past broadcasts of each broadcaster every day. Because every past broadcast only counts the number of viewers during its broadcast, the number of total views indeed reflects the characteristics of live streams. Table~\ref{tab:api} shows the details of the REST APIs used in our crawler. Our dataset includes $2,923$ active broadcasters (i.e., sources), who have broadcast a total of $105,117$ live performances, attracting over $17.8$ million viewers. That is, each source has conducted around $36$ live broadcasts in the two-month period. These broadcasts are of different durations and viewer populations, as we will analyze in the next section.
%XXX
\begin{figure}[!t]
  \center
  \includegraphics[width=0.4\textwidth]{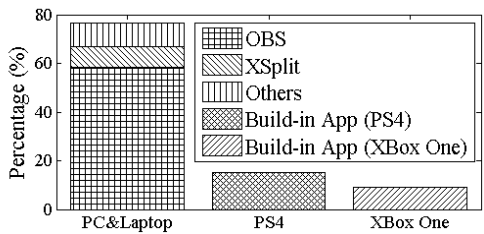}
  \caption{Device distribution of Twitch's live sources}
  \label{fig:platform}
\end{figure}

The broadcast sources can be quite heterogeneous, involving PCs, laptops, and even PS4/XBox game consoles, and multiple sources can be involved in one broadcast event. For instance, recent DotA2 (Defense of the Ancients 2) game championships ``The Summit 2" embraces at least three sources to stream this event, including two game competitive players and a commentator's perspective.
Figure~\ref{fig:platform} plots the distribution of the broadcasters' devices in Twitch. Given that the build-in Apps of PS4/Xbox were available just after March 2014, we can clearly see that the PC/Laptop are the most popular devices, at about $76.7\%$; the second is PS4, at about $15.1\%$; and the third is XBox One, at about $9.2\%$. This figure also indicates that the most widely used streaming software on PC/laptop platform is Open Broadcaster Software (OBS)~\footnote{https://obsproject.com}, at about $59\%$.

Our analysis results show that Twitch deploys RTMP (Real Time Messaging Protocol over HTTP Tunnel) streaming servers, covering 14 regions, to compensate the weaknesses of the sources, e.g., networking fluctuation and inferior performance.
The original streaming will be transferred through HTTP Live Streaming from streaming servers to viewers with the assistance of a CDN, whereas all the servers are of names: video\#.sfo\#\#.hls.twitch.tv \footnote{The name also indicates the location of the corresponding CDN server; e.g., ``sfo" for San Francisco.}.
It is known that Twitch further deploys load balancing servers (usher.twitch.tv) to optimize the live streaming distribution~\cite{Twitch:justin} and deliver HTTP Live Streaming playlist file {\em channelID.m3u8} to each viewer's device. To accommodate heterogeneous viewers, Twitch also provides adaptive online transcoding service to premium content broadcasters. All the live performances can be watched by web browsers or Twitch Apps for mobile devices (e.g., iOS or Android-based).
If a premium broadcaster enables online transcoding service, the browser-based viewers can manually select a proper quality from~\texttt{Source},~\texttt{High},~\texttt{Medium},~\texttt{Low}, and~\texttt{Mobile}, and the option ~\texttt{Auto} (i.e., adaptive streaming) is the default setting for a mobile user. However, as we will show, the duration of $50\%$ sessions are over $150$ minutes, which imposes too much overhead to transcoding, and hence non-premium broadcasters can only make a tradeoff by selecting a streaming quality for most of the viewers.

Interactive communication is a unique feature in such a Twitch-like crowdsourced system. A set of interactive messaging servers receive the viewer's live messages, and then dispatch the messages to the corresponding live broadcaster and other viewers, enhancing the participants' experience for the live events towards realistic competition environment. That said, the viewers are no longer passive, but can affect the progress of the broadcast as well. In particular, for broadcasting live game playing, the interactive service allows viewers to interact with the game players and commentators in realtime. Our data reveal that these servers for interaction are only deployed in North America using the IRC (Internet Relay Chat) protocol; yet they deliver all the live messages worldwide with reasonably low latency, as we will show in Section~\ref{sec:livemsg}.

\begin{figure}[!t]
  \center
  \includegraphics[width=0.45\textwidth]{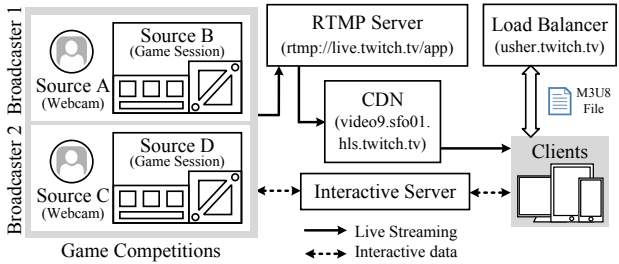}
  \caption{Two broadcaster/game player measurement configuration}
  \label{fig:event}
\end{figure}

To closely investigate the behavior and experience of individual sources and viewers, we also set up three source-end PCs (one commentator and two game players) and five viewers over the Twitch platform, and use network tools, including Wireshark, tcpdump, and ISP lookup, to monitor their detailed incoming and outgoing traffic. Figure~\ref{fig:event} describes the basic two-player competition broadcast setup for game DotA2. Each player has installed a web camera that captures the video in realtime and encodes in H.264 locally with OBS v0.63b, which is then transmitted to the Twitch platform through RTMP. All devices in our platform are of household PC/tablet configurations, which ensure that our measurement results are representative for general users. The configuration of each device is shown in Table~\ref{tab:broadcaster:device} and \ref{tab:viewer:device}. The iOS and Android devices were jail-broken/rooted to capture the incoming/outcoming traffic precisely. We also deployed a NETGEAR GS108PEv2 switch to simulate the dynamic uploading bandwidth on the hardware level, which is much more accurate than a software limiter. Finally, to quantify the latencies on the viewer's side and the impact of network dynamics on Quality-of-Experience (QoE), we use the commentator's laptop (B1) as NTP (Network Time Protocol) server and synchronize other devices to improve the accuracy of measurement results.

\begin{table}[!th]
\small
\center
\caption{The configuration of broadcasters' device (B1: Commentator; B2, B3: Players)}
\begin{tabular}{|l|l|l|l|l|}
\hline
\textbf{ID} & \textbf{Type} & \textbf{Operating Sys.} & \textbf{Uploading}\\ \hline
B1               & Laptop              & Windows 8.2                 & 2-12 Mb/s\\ \hline
B2               & Desktop             & Windows 8.2                 & 5-18 Mb/s\\ \hline
B3               & Desktop             & Windows 7                   & 3-15 Mb/s\\ \hline
\end{tabular}
\label{tab:broadcaster:device}
\end{table}
\begin{table}[!th]
\small
\center
\caption{The configuration of viewers' device}
\begin{tabular}{|l|l|l|l|l|}
\hline
\textbf{ID} & \textbf{Network} & \textbf{Operating Sys.} & \textbf{Downloading}\\ \hline
P1               & Wired                 & Windows 7                & 160-250 Mb/s\\ \hline
P2               & Wireless              & Windows 8.2                & 7-25 Mb/s\\ \hline
M1               & Wireless              & iOS 8.0                    & 2-25 Mb/s\\ \hline
M2               & Wireless              & Android 4.2.2              & 4-36 Mb/s\\ \hline
M3               & 3G                    & Android 4.2.2              & 0.6-1.2 Mb/s\\ \hline
\end{tabular}
\label{tab:viewer:device}
\end{table}
\section{View Statistics and Patterns}
\label{sec:viewfeatures}

We analyze Twitch views data and find that it represents several novel and unique characteristics. As of October, 2014, the peak of concurrent streams is above 12000, most of which are for online gaming broadcast. These game streams attract more than one million views every day. We first investigate the characteristic of views in different live contents, and then discuss the source-driven and event-driven views.

\subsection{Popularity and Duration}
\begin{figure}[!t]
\begin{minipage}[b]{0.9\linewidth}
  \center
  \includegraphics[width=1\textwidth]{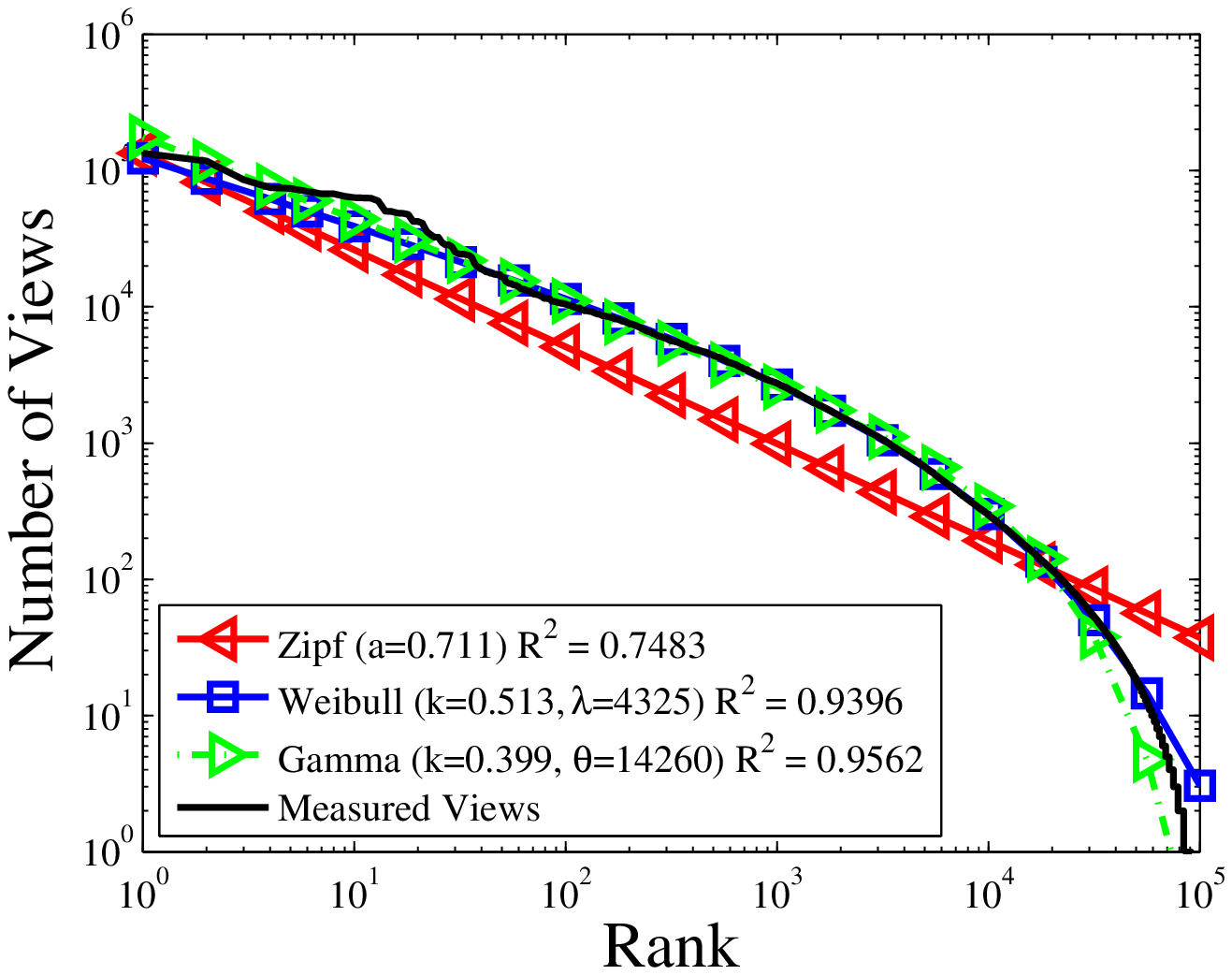}
  \caption{Live streams rank ordered by views}
  \label{fig:vvss}
\end{minipage}
\\
\begin{minipage}[b]{0.9\linewidth}
  \center
  \includegraphics[width=1\textwidth]{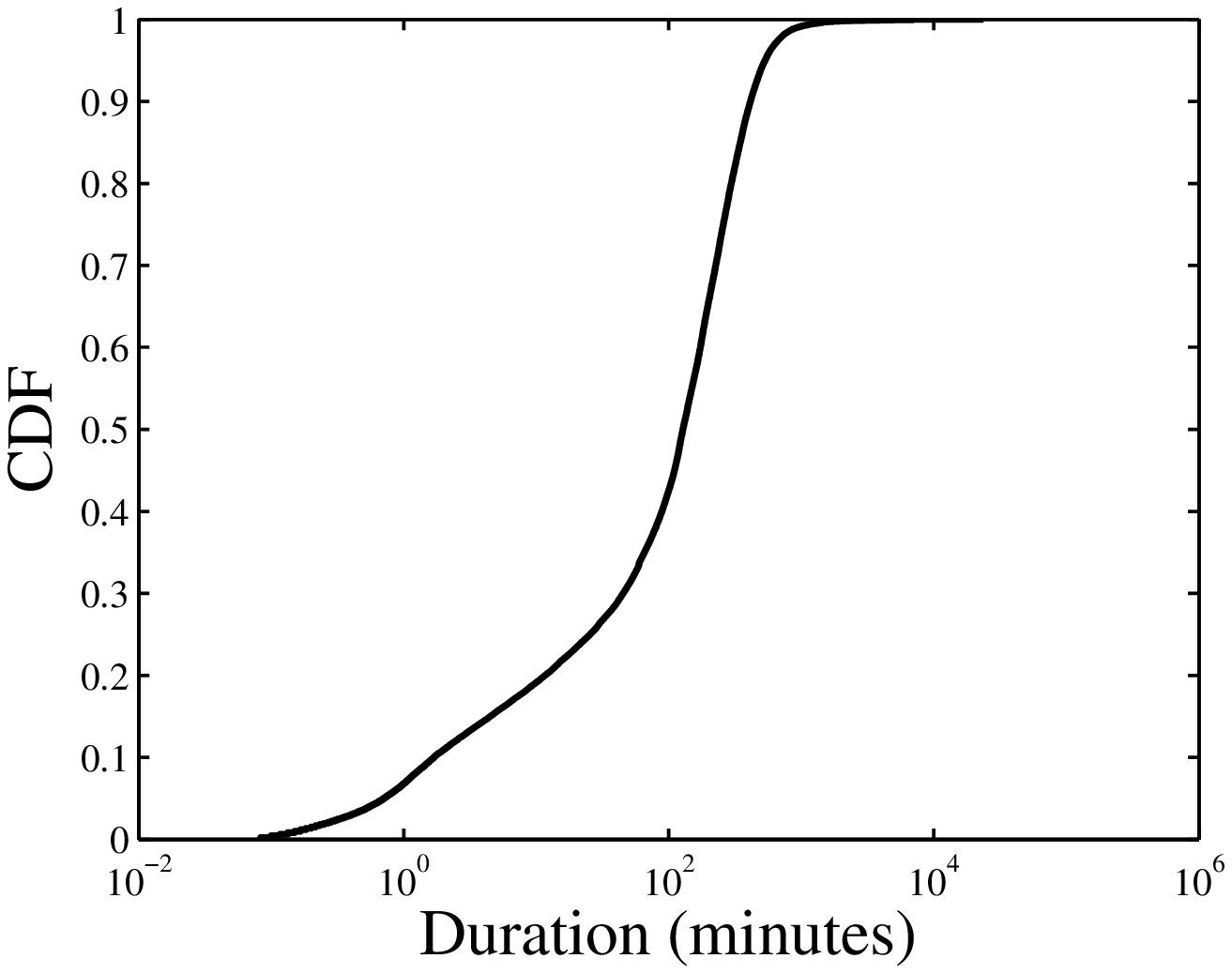}
  \caption{Distribution of live streaming duration}
  \label{fig:duartion}
\end{minipage}
\end{figure}

\begin{figure}[!t]
\center
\subfloat[Total views]{
\includegraphics[width=0.45\textwidth]{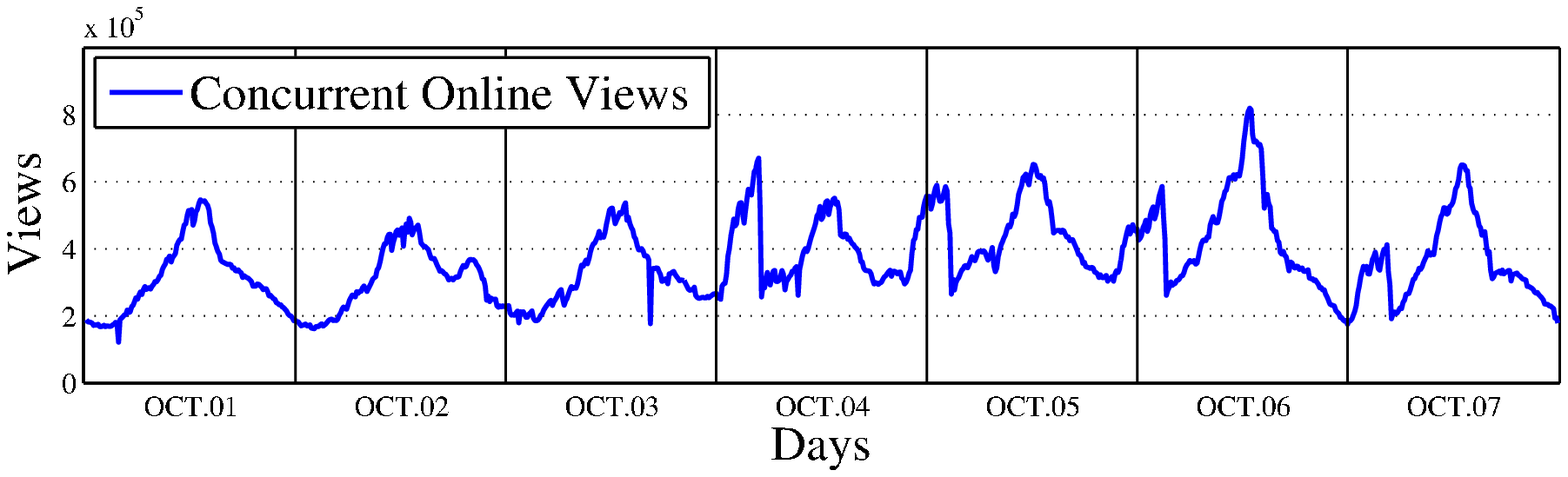}
\label{fig:totalviews}}
\\
\subfloat[League of Legends]{
\includegraphics[width=0.45\textwidth]{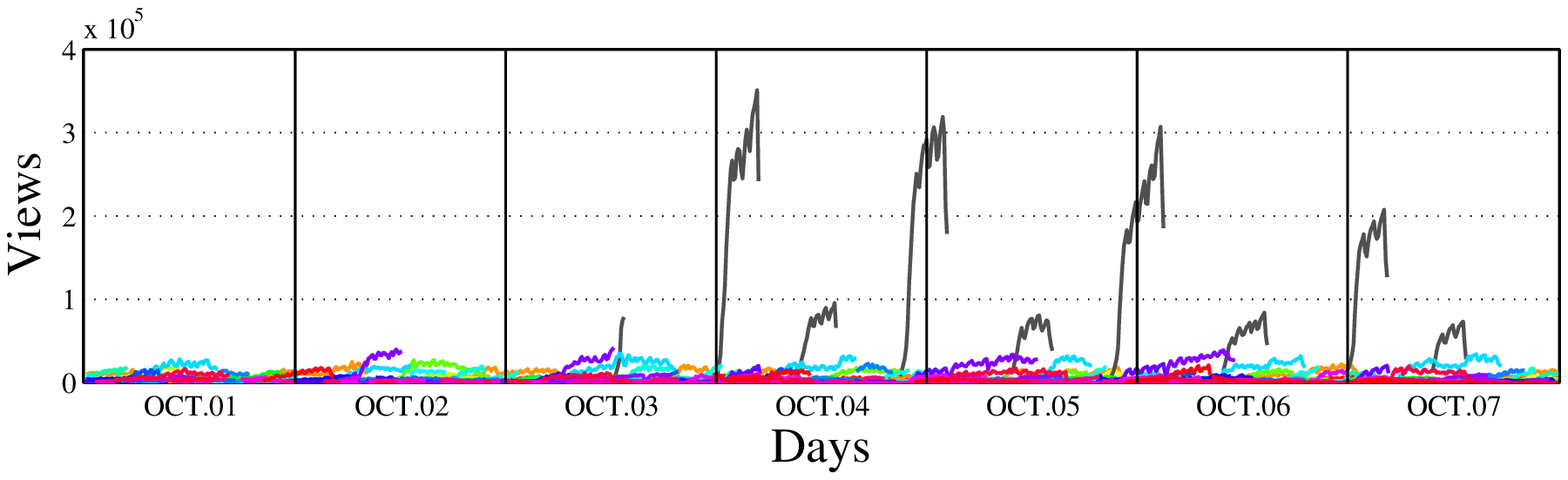}
\label{fig:lolviews}}
\\
\subfloat[Defense of the Ancients 2]{
\includegraphics[width=0.45\textwidth]{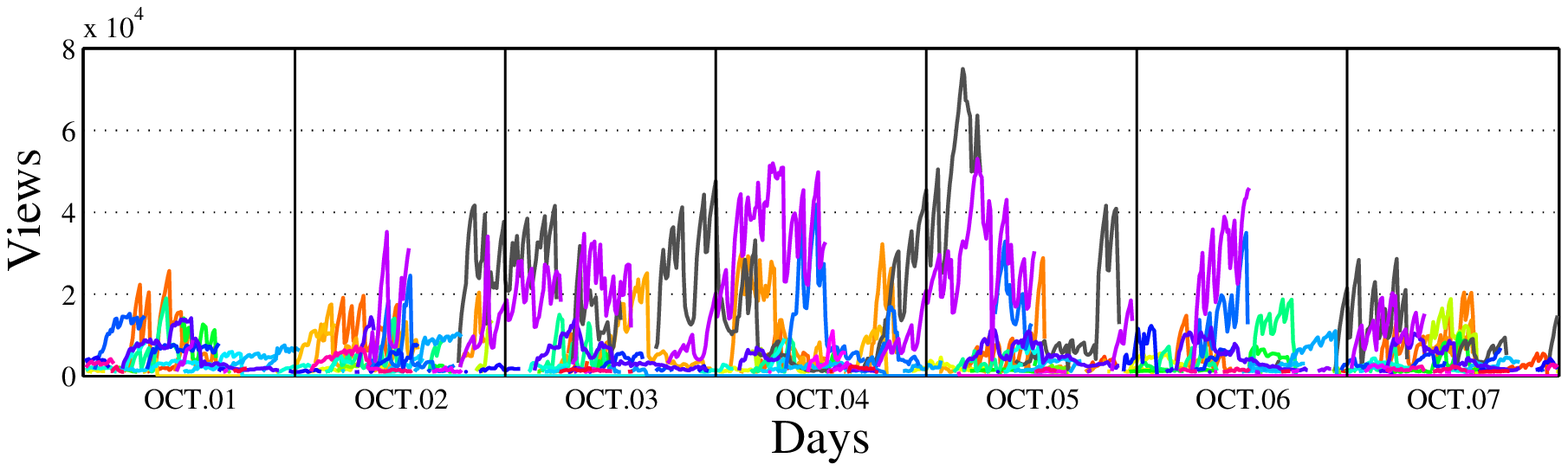}
\label{fig:dotaviews}}
\caption{Views patterns in Twitch (From 2014OCT01 to 2014OCT07)}
\label{fig:views}
\end{figure}

The number of viewers is one of the most important characteristics, which reveals the popularity and access patterns of the content. For our global view dataset containing more than 105 thousand streams, we plot the number of views as a function of the rank of the video streams' popularity in Figure~\ref{fig:vvss}. Clearly, the plot has a long tail on the linear scale; it does not fit a classical Zipf distribution, which is a straight line in a log-log scale, as shown in Figure~\ref{fig:vvss}. We also plot two other typical distributions, Weibull and Gamma. Because they have heavy tail, especially in the top part, and have been demonstrated to be better fits in YouTube~\cite{Xu:2013:YouTube}, they are also good in the Twitch's case, either. We also calculate the coefficient of determination $R^{2}$ to indicate the fitness in this figure. Weibull and Gamma distributions can fit the rise part, in which the popular streams hosted by famous players or commentators attract a large number of game fans through broadcasting game competitions. We also analyze the influences of live events in Section~\ref{sec:eventviews}.

To understand Twitch's uniqueness, we closely examine the relationship among the total number of views and broadcasters, and the number of views in top broadcasters every ten minutes in our dataset. We find that top-$0.5\%$ broadcasters contribute to more than $70\%$ of the total views in general.
In several extreme cases, the top-$0.4\%$ broadcasters account for more than $90\%$ of total views. As such, the distribution of the views in Twitch exhibits extreme skewness, being much stronger than conventional streaming systems, e.g., YouTube~\cite{Zink:2009:CN} and PPLive~\cite{Li:2012:IMC:PPLive}.
We will illustrate the reasons of this huge skewness in next subsection.

When considering the computation and traffic impacts of a broadcast, popularity is not the only factor, which must be weighted together with the duration of a broadcast. As shown in Figure~\ref{fig:duartion}, the streaming durations are highly diverse, too.
About $30\%$ live contents have a duration around $60-120$ minutes, but there are also $30\%$ being more than $4$ hours, which is dramatically longer than those in typical user-generated video sharing platforms, e.g., YouTube, where the longest steam is around 2-3 hours (i.e., movies)~\cite{Xu:2013:YouTube}. The exact duration of a Twitch broadcast, which depends on the interest of and the interaction with the viewers, can hardly be predicted in advance, either. This again is different from professional content services, e.g., TV broadcast. Such long-lived and yet unpredictable broadcast apparently pose challenges on computation and bandwidth resource allocation for real-time online transcoding and reliable broadcasting.

\subsection{Event- and Source-Driven Views}
\label{sec:eventviews}
Due to the globalized demands with time/region diversities, it is well-known video services always experience dynamics and fluctuations requests~\cite{Fei:2015:INFOCOM}. To understand the view dynamics of Twitch, Figure~\ref{fig:totalviews} depicts the online views over time in a one-week period (from OCT01 to OCT07, 2014). The number of concurrent online views exhibits daily patterns: like in the conventional video services~\cite{Zink:2009:CN}, the Twitch viewers tend to watch game streaming during the day and evening, whereas less likely in midnight. Interestingly, the number of views was the highest around the midnight on OCT04 and then hastily decreased to the lowest level, implying that if an prominent source can indeed attract massive viewers, despite of time. Similar (though less striking) patterns can be seen in OCT05, 06, and 07.

 There are also two transient drops from time to time, e.g., on OCT03. After investigating the broadcasters' data, we find that a popular live streaming was disconnected for an unknown reason but re-connected quickly. Accordingly, the number of viewers decreased instantly but managed to recover in a few minutes after re-connection. Such situations rarely happen for professional broadcasters, which have highly reliable equipment setup and network connections. Crowdsourced broadcast system, e.g., Twitch, on the other hand, relies on the non-professionals to provide the broadcast content in realtime; as such, even if the Twitch platform itself is highly reliable with over-provisioned resources, it can hardly guarantee the source video quality.

 To further understand the roles of the sources, Figure~\ref{fig:lolviews} and \ref{fig:dotaviews} detail the number of views among top broadcasters in two game categories (League of Legends, and Defense of the Ancients 2) during one week. As can be seen, the broadcast can be suspended suddenly; e.g., there are four obvious rises in Figure~\ref{fig:lolviews} which dropped immediately, due to terminating the game competitions. Since the live progress depends on what is actually happening in the game competitions, the duration and the exact time of termination can hardly be predicted (see for example the variations in~\ref{fig:dotaviews}). The exact reason and time that trigger a views burst can hardly be predicted, either. Note that these still hold with content other than gaming, as long as they are provided by distributed crowdsources. In short, the views of a crowdsource live streaming system can be more dynamic and unpredictable than conventional video services, and the views are both {\em event-} and {\em source-driven}. Even though the Twitch platform is aware of the online status of the massive sources and viewers, significant efforts are still needed to provide persistently good user experience.

\section{Messaging and View Latency}
\label{sec:viewer}
We next examine the latencies in the Twitch system, which are critical to the user experience, particularly with live interactions. To this end, we focus on
the latencies experienced by a set of representative viewers with typical device and network settings, namely, wired PC viewer (P1), wireless PC viewer (P2), and mobile tablet viewers (M1, M2, M3). Three latencies are of interest here, namely, {\em live messaging latency}, {\em broadcast latency}, and {\em switching latency}.

\subsection{Live Messaging Latency}
\label{sec:livemsg}
\begin{figure}[!t]
\center
\subfloat[Live messaging latency]{
\includegraphics[width=0.23\textwidth]{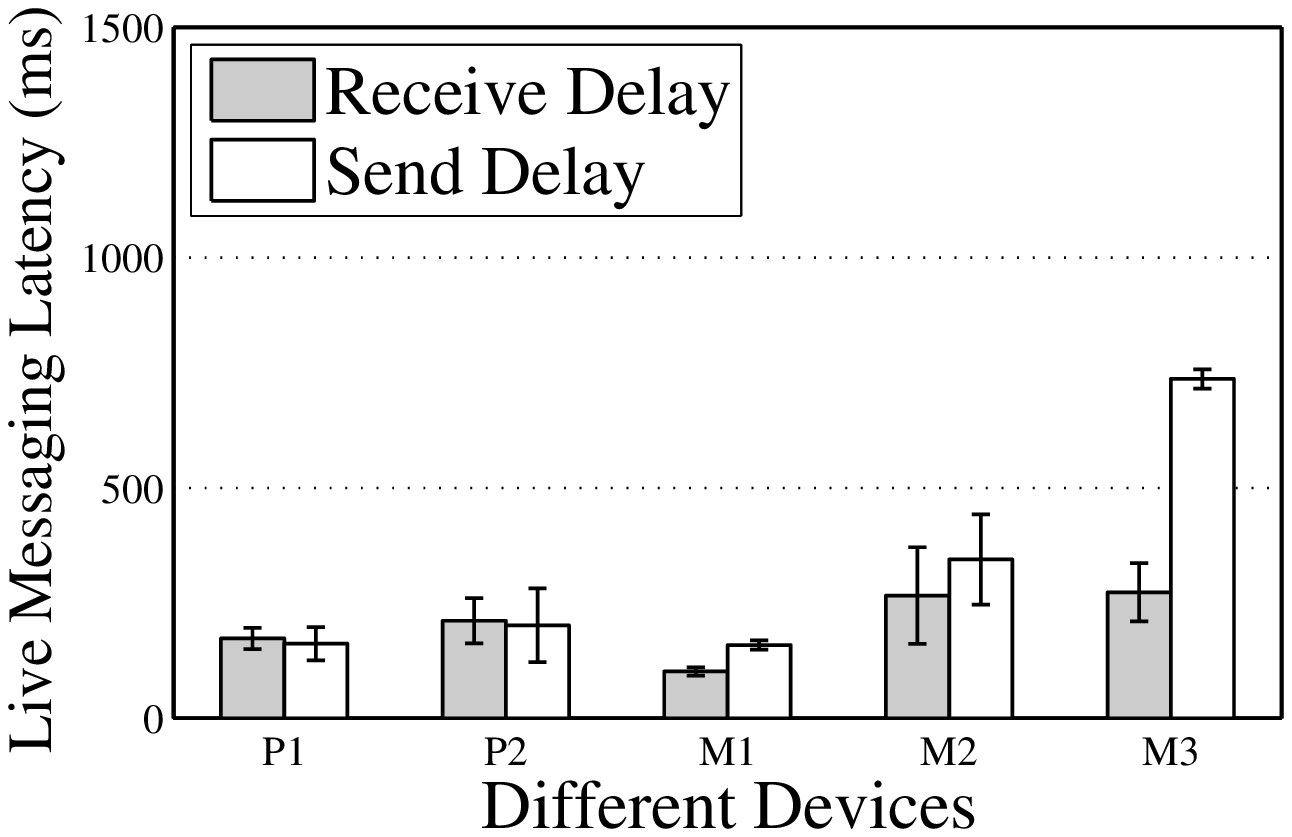}
\label{fig:LiveMessageLatency}}
\hspace{-2mm}
\subfloat[Broadcast delay]{
\includegraphics[width=0.23\textwidth]{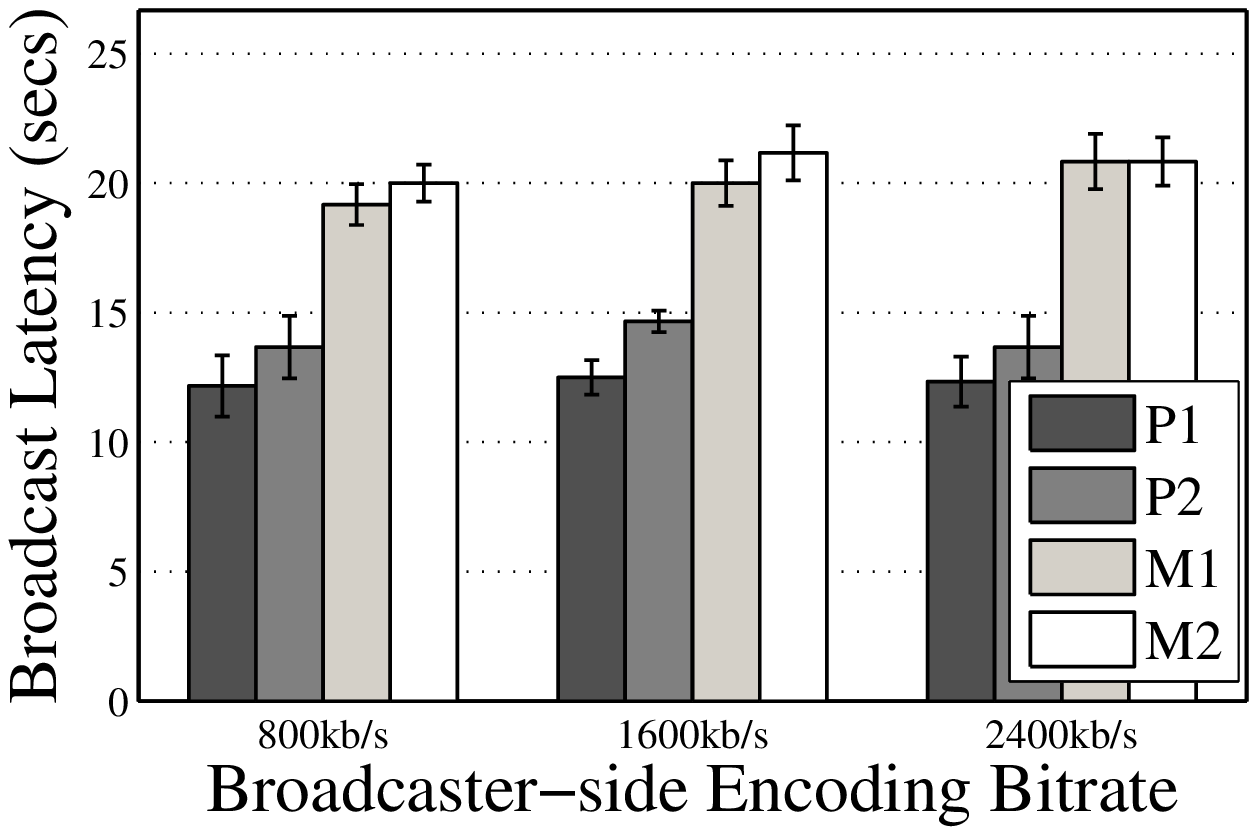}
\label{fig:latency}}
\hspace{-2mm}
\subfloat[The impacts of networks]{
\includegraphics[width=0.23\textwidth]{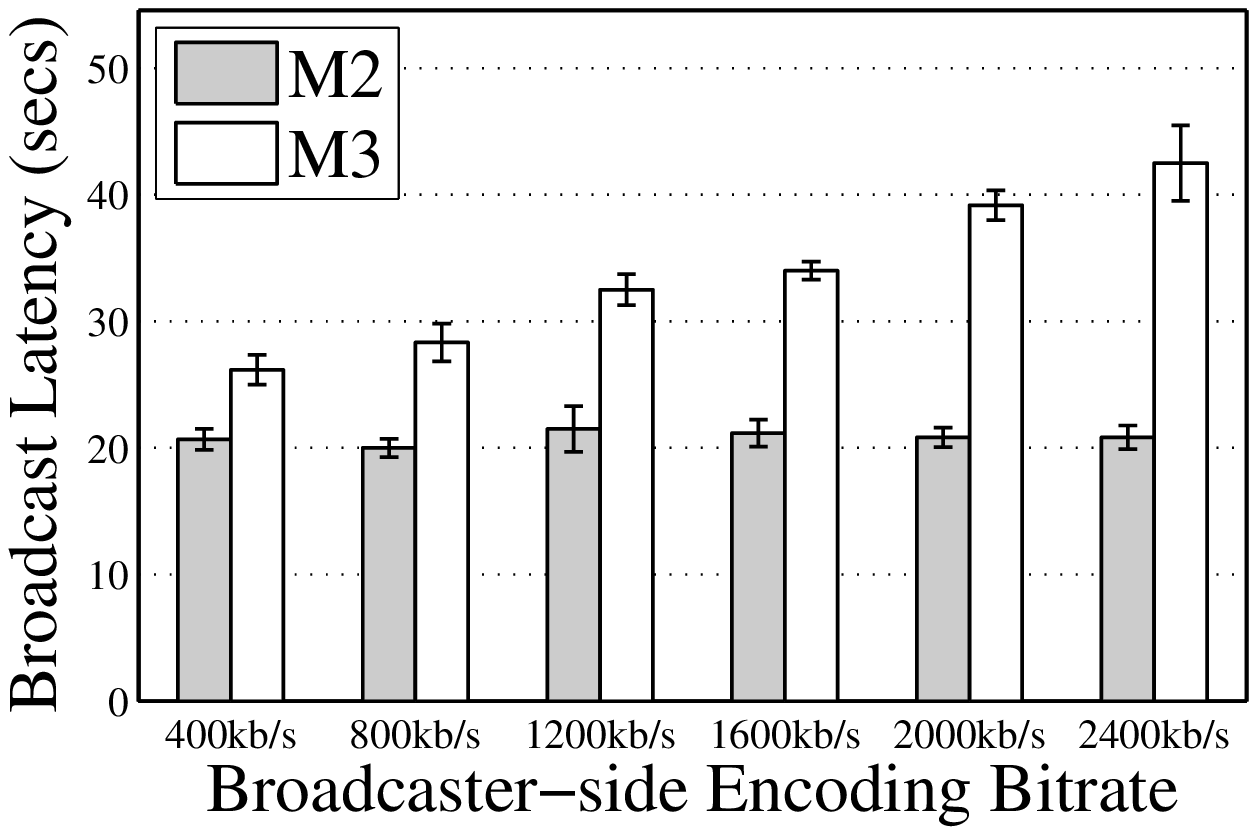}
\label{fig:networking}}
\hspace{-2mm}
\subfloat[Source switching latency]{
\includegraphics[width=0.23\textwidth]{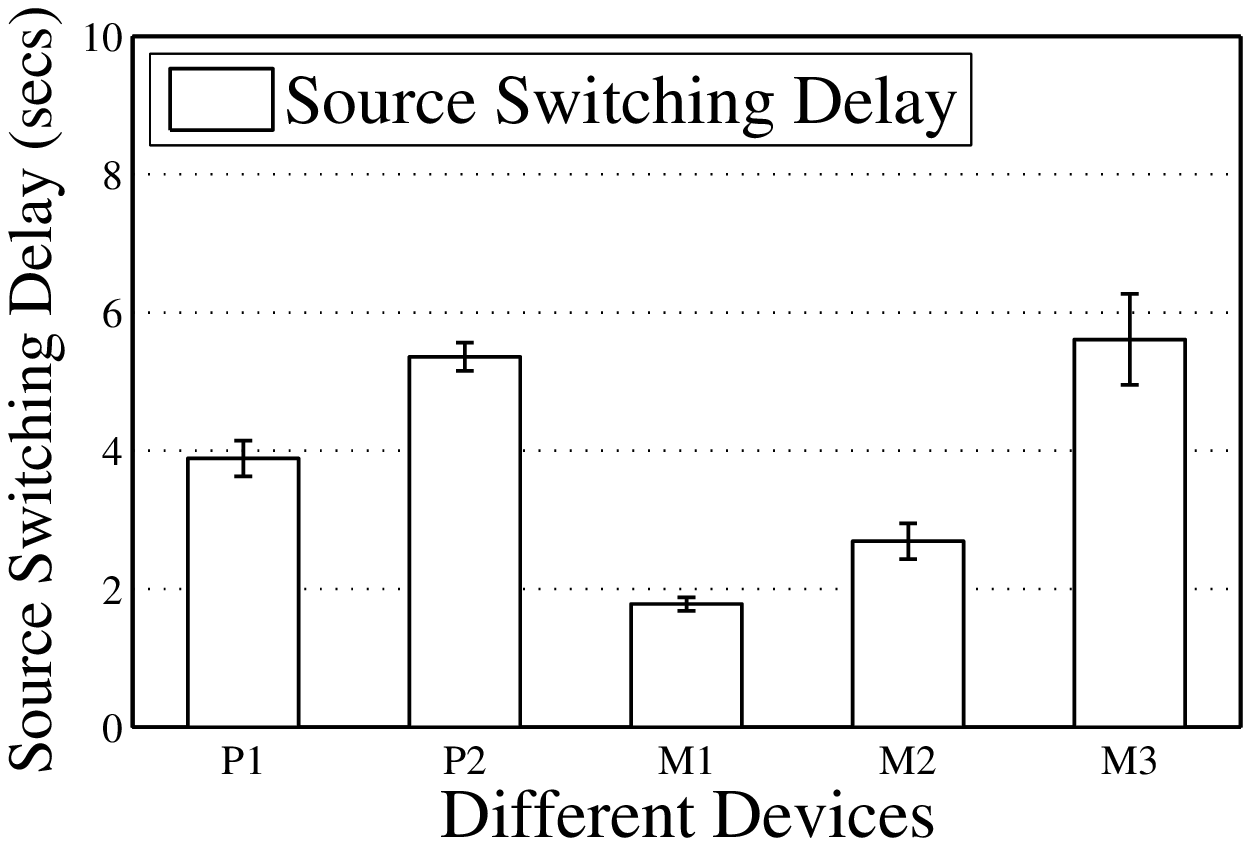}
\label{fig:SSLatency}}
\caption{The characteristics at the viewer-side (error bars are 95\% confidence intervals)}
\label{fig:twochannels}
\end{figure}

A distinct feature of the crowdsourced content production is that all viewers and broadcasters can interact and discuss the current live events, which collectively affect the ongoing and upcoming broadcast content. Twitch enables the collaboration via {\em live messages} exchanged through a set of interactive servers, as shown in Figure~\ref{fig:event}. We capture the networking traffic from five devices and three sources and analyze the send/receive timestamps of live messages (see Section 2 for the experiment configuration). Figure~\ref{fig:LiveMessageLatency} presents the live message latencies for five representative viewer devices in our experiments. This type of latency depends on both network conditions and device types; two desktop devices witness the almost same latency between the message sending and receiving operations, whereas the receiving latency of mobile devices is lower than the sending latency. Yet the measurement results suggest that, in general, the live message is quite low ($\leq 400ms$), enabling responsive interaction among the participants (viewers and broadcasters). It is worth noting that, along with the live message, certain control information including a participant's type and streaming quality is also sent to a Twitch statistic server (mp.twitch.tv), as found in our captured network data.

\subsection{Broadcast Latency}
We next measure the {\it broadcast latency}, which is defined as the time lag of a live event when viewers watch the live streaming from the source. It reflects a viewer's time difference with the commentator and other viewers when they watch and discuss the current live event. A long broadcast latency will obviously affect the interactivity.

Figure~\ref{fig:latency} shows the average, maximum, and minimum broadcast latencies of the four viewer devices (P1, wired PC; P2, wireless PC; M1, M2, mobile tablet). We first vary the streaming bitrates from 800 Kb/s to 2400 Kb/s, and ensure that the downloading bandwidth of each device is significantly higher than the streaming bitrate, so as to mitigate the bottleneck within the network. As shown in the figure, the browser-based P1 and P2 have a latency about 12 seconds for different streaming rates, whereas the client-based M1 and M2 have about 21 seconds. We closely investigate the traffic of each device and find that Twitch adopts a device-dependent broadcast latency strategy to gather the crowdsourced content for processing and to ensure smoothed live streaming.
For desktop devices, the inevitable latency derives from that Twitch receives and converts RTMP streaming to HTTP Live Streaming chunks, each of which is a four-second streaming segment; on the other hand, for mobile devices, Twitch will strategically send three more chunks than desktop devices, if all devices start to play live streaming simultaneously. That is, even if we consider an ideal network, mobile devices will still suffer an extra 12 seconds broadcast latency in the current Twitch platform.
To evaluate the impact of network bottlenecks, we also compare the latencies of mobile devices with WiFi (M2) and 3G (M3) networks, as shown in Figure~\ref{fig:networking}. As can be seen, the latencies for M2 remain almost constant across all streaming rates, and M3 incurs extra network delays that increase with the growing streaming rate. The extra network latency is significant only with very high streaming rates,
which implies that the processing time (about 10s for all devices) and strategic delivery (extra three chunks for mobile devices) within the Twitch platform are the key factors in broadcast latency.

\subsection{Source Switching Latency}
Given the massive sources available, a viewer has rich choices and can frequently switch among different sources, for the same broadcast event, or even to a totally different event, both of which are now done manually in Twitch (per viewer's action). To investigate the latency of source switching, we record the time duration for $100$ switches performed by the different types of devices in different network environments, as shown in Figure~\ref{fig:SSLatency}. Not surprisingly, a higher downloading bandwidth enables a lower switching latency in both wired and wireless networks (e.g., 4 seconds for a high speed wired network and 5.5 seconds for a low speed wireless network). The latency however is not proportional to the bandwidth; in particularly, the devices in the mobile networks generally have lower switching latencies than those in the wired network, although the mobile bandwidths are indeed much lower, which again indicates different device-dependent strategies have been applied within Twitch.

\subsection{Impacts of Broadcaster's sources}
\label{sec:broadcaster}
So far, we have examined the latencies on the viewer's side, which includes not only the processing time within the Twitch server and the time from the server to the viewer, as in conventional streaming systems, but also the latency from the source to the server, a new component in the crowdsourced system. Through household Internet accesses and multimedia-ready PCs or mobile devices, anyone can become a Twitch broadcaster, anywhere and anytime. These non-professional broadcasters however have diverse networking connections, both in terms of capacity and stability, especially with wireless mobile accesses. To evaluate the network impact, we deploy a modified OBS module on every broadcaster to record the bandwidth consumption, and first initialize live streaming service in the networks with sufficient uploading bandwidth.

\begin{figure}[!t]
\center
\includegraphics[width=0.5\textwidth]{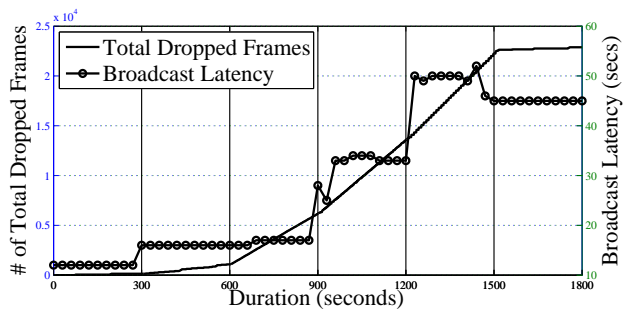}
\caption{The impacts of broadcaster's network}
\label{fig:delayframe}
\end{figure}

To understand the impact, we next control the maximum uploading bandwidth following five settings: No Limit, 4000 Kb/s, 2000 Kb/s, 1000 Kb/s, and 512 Kb/s; each one lasts five minutes (300 seconds), and the setting finally returns to No Limit at the 1500 second. The original streaming encoding setting is still 4000 Kb/s, and the measurement results are shown in Figure~\ref{fig:delayframe}. From this figure\footnote{For simplicity, we only show the broadcast latency between P1 and B1. To avoid measurement bias, we repeat the same test on another two broadcasters' devices B2/B3 and other viewer's devices. The results remain consistent with Figure~\ref{fig:delayframe}.}, we observe that the number of total dropped frames consistently grows with decreasing the uploading bandwidth on the broadcaster's side. In the meantime, the broadcast latency on the viewer's side also suffers the stepwise rise; in particular, the live streaming experiences two notable delay increases at 900 and 1200 seconds. That said, Twitch attempts to maintain a stable broadcast latency, but cannot guarantee the smooth live streaming. Another interesting phenomenon occurs after recovering the broadcaster's uploading condition (1500-1800 seconds). In this case, the uploading capacity becomes sufficient again, and the broadcaster can offer a stable streaming to Twitch; yet Twitch just decreases the broadcast delay slightly to mitigate the impacts of previous networking diversity at the broadcaster-side. These measurement results indicate that the streaming service provided by Twitch is vulnerable and sensitive when the broadcaster's networking capacity is changed frequently, not to mention responsive interactions.

\section{Conclusion and Further Discussion}
\label{sec:conclusion&further}
Multi-sourced live streaming from non-professional broadcasters have emerged in the market, and is rapidly evolving to crowdsourced with massive participants. In just two years, the most successfully platform, Twitch.tv, has grown to be 4th largest traffic generator in the US Internet, and is with a steady $8\%$ monthly growth rate now. In this paper, we presented an initial investigation on the modern crowdsourced live streaming systems,
using Twitch as a case study. Closely examining the access data collected in a two-month period, we
outlined the inside architecture of Twitch, and revealed that the views patterns are determined by both the event and the broadcasters' sources. Our
measurement also explored the unique source-driven and event-driven views, showing that the current delay strategy on the viewer's side substantially impacts the viewers' QoE, and there is significant inconsistency among the long broadcast latency and the short live messaging latency. On the broadcaster's side, the dynamic uploading capacity is a critical challenge, which noticeably affects the smoothness of live streaming for viewers.

As a preliminary study, both the scale and the interactions we have considered are limited. The Twitch-like services themselves remain in the infancy stage, too. In February 2014, a pilot project ``Twitch Plays Pokemon" offered live streaming and game emulator for the game Pokemon Red, in which players (also as the viewers in Twitch) simultaneously send the control message of Pokemon through the IRC protocol and live messages in Twitch. This truly crowdsourced game streaming attracted more than 1.6 million players and 55 million viewers. Similar scales however have yet to appear in other interactive events, though. It is also known that the user interaction experience is not very satisfied in the pilot project, which is due largely to the latency disparity between live messages and the broadcast content, as we have quantified through measurement.

Joint optimization of servers and clients has been commonly employed in state-of-the-art streaming services to provide smooth streaming experience for heterogeneous viewers\cite{Zhi:2014:INFOCOM}. For Twitch-like services, it is necessary to include the massive sources in the optimization loop, which however can be quite challenging given their strong dynamics. Yet the crowdsourced nature provides opportunities, too. We suggest that, through analyzing the enormous amount of historical activities of the broadcasters and viewers, the service provider may predict their behaviors in advance and accordingly improve the streaming quality.

Our measurement indicates that the adaptation strategy in Twitch is mainly based on CBR video. As such, good smoothness and low latency can hardly be both offered with limited bandwidth. Even though viewers can potential switch from a high-latency source to a low-latency one, a certain period of live streaming would be missed given the high switching latency. For the viewer side, existing work~\cite{Pires:2014:VideoNext} proposed a trade-off solution to distribute adaptive streaming in Twitch, which allows the heterogeneous viewer's devices to adaptively select the best-fit streaming quality.
For the source side, we are working on a crowd uploading strategy that attempts to leverage the aggregated bandwidth of the many sources for speedy uploading. The live messaging and the associated social connections can play useful roles in the uploading, too.

\bibliographystyle{IEEEtran}
\bibliography{twitchref}
\end{document}